\documentclass[journal]{IEEEtran}
\ifCLASSINFOpdf
\else
\fi
\usepackage{graphicx}
\usepackage{amssymb}       
\usepackage{amsmath}
\usepackage{cases}
\usepackage{epstopdf}
\usepackage{cite}
\usepackage{fancyhdr}

\begin{document}
%
\title{Online Reinforcement Learning Control by Direct Heuristic Dynamic Programming: from Time-Driven to Event-Driven}
%
%
%

\author{Qingtao~Zhao, Jennie~Si,~\IEEEmembership{Fellow,~IEEE,} and~Jian~Sun,~\IEEEmembership{Member,~IEEE}
	\thanks{Q. Zhao and J. Sun are with the Key Laboratory of Intelligent
		Control and Decision of Complex System, Beijing Institute of Technology, Beijing, 100081, China (e-mail:zhaoqingtaophx@163.com; sunjian@bit.edu.cn).}
	\thanks{J. Si is with the Department of Electrical, Computer,
		and Energy Engineering, Arizona State University, Tempe, AZ 85287 USA
		(e-mail: si@asu.edu). The research of J. Si is supported in part by NSF \#1563921 and \#1808752.}
	}

%
%

\markboth{Journal of \LaTeX\ Class Files,~Vol.~14, No.~8, August~2015}%
{Shell \MakeLowercase{\textit{et al.}}: Bare Demo of IEEEtran.cls for IEEE Journals}
%



\maketitle

\begin{abstract}
In this paper time-driven learning refers to the machine learning method that updates parameters in a prediction model continuously as new data arrives. Among existing approximate dynamic programming (ADP) and reinforcement learning (RL) algorithms, the direct heuristic dynamic programming (dHDP) has been shown an effective tool as demonstrated in solving several complex learning control problems. It continuously updates the control policy and the critic as system states continuously evolve. It is therefore desirable to prevent the time-driven dHDP from updating due to insignificant system event such as noise. Toward this goal, we propose a new event-driven dHDP. By constructing a Lyapunov function candidate, we prove the uniformly ultimately boundedness (UUB) of the system states and the weights in the critic and the control policy networks. Consequently we show the approximate control and cost-to-go function approaching Bellman optimality within a finite bound. We also illustrate how the event-driven dHDP algorithm works in comparison to the original time-driven dHDP.

\end{abstract}


%
\IEEEpeerreviewmaketitle

\section{Introduction}
%
%
%
%

\IEEEPARstart{I}{n} this paper time-driven learning refers to the machine learning method that updates parameters in a prediction model continuously as new data arrives. While important applications such as autonomous robots and intelligent agents on the web or in mobile devices require on-the-fly and continuous adaptation to environment changes, catastrophic forgetting \cite{mccloskey-1989,Ratcliff-1990} can occur and thus drastically disrupt prediction performance. From a human learning perspective, such periodic, continuous update of the prediction model does not best represent biological learning. Long established theory based on extensive experimental data shows that short-term memory decays very slowly when the environment is undisturbed, but decays amazingly fast otherwise. In machine learning, disruption of a learned model by new learning is a recognized feature of neural networks \cite{Carpenter-1986,Hinton-1984,Hinton-1987,Ratcliff-1990,Sutton-1986,Li-2017,Zenke-2017}.

Forgetting after learning has plagued the machine learning field for many years and it has attracted attention from machine learning researchers who seek scalable and effective learning methods. A notable such progress was reported in \cite{Kirkpatrick-2017} where several Atari 2600 games were learned sequentially by elastic weight consolidation (EWC) algorithm. To learn a new task, EWC tempers the network parameters based on previous task(s): it enables fast learning rates on parameters that are poorly constrained by the previous tasks and slow learning rates for those that are crucial. Incremental learning is an idea of triggering learning only if new patterns are sufficiently different from previous ones \cite{Carpenter-1992,Polikar-2001}. For example, the Learn++ algorithm is to make the distribution update rule contingent on the ensemble error instead of the previous hypothesis error to allow for efficient incremental learning of new data that may introduce new classes. In \cite{Li-2017}, learning without forgetting (LwF) was proposed by employing convolutional neural networks. Given a set of shared weights across all tasks, it optimizes the shared weights, old task weights and new task weights simultaneously based on an objective function including the predicted errors of old tasks, new tasks and a regulation term. Similarly, to alleviate catastrophic forgetting, Zenke $et \ al.$ allowed individual synapses to estimate their importance for learned task. When the weights learn a new task, the cost-to-go function is modified by adding a penalty term of the summed parameter error functions of all previous tasks to reduce large changes in important parameters \cite{Zenke-2017}.

While the above discussions are pertinent to supervised or unsupervised learning, forgetting after learning can also plague RL or ADP. The problem has received some attention. In \cite{Zhong2-2016} and \cite{Yang-2017}, the authors established a framework to update neural network weights in an event-triggered manner while the weight convergence and system stability were still guaranteed.
But the results were obtained for nonlinear affine
systems. Additionally, an observer network was required in \cite{Zhong2-2016} and a system identifier network was required in \cite{Yang-2017}. In \cite{Dong-2017,Zhu-2017,Zhang-2019}, the authors considered a partially unknown affine nonlinear system. They showed the weight convergence of the neural
networks and controlled system stability while taking into account control input constraints and under different event triggering conditions. Also dealing with nonlinear affine systems, \cite{Zhang-2016,Wang-2016,Wang-2017,Zhang2-2016,Yang-2018} take into account an internal uncertainty or an external disturbance by the proposed event-triggered robust control using ADP structures. A different event triggering condition based on value functions was proposed in \cite{Luo-2019}  for continuous-time nonlinear systems using ADP. The convergence of the performance index was guaranteed while the system stability and weight boundedness were also considered. But such triggering condition is indirect and its dependence on value function may result in false triggers as value function approximation errors vary, sometimes to a large degree.

Among different ADP algorithms \cite{Si-2004,Lewis-2013}, heuristic dynamic programming (HDP) provides a basic structural framework of adaptive critic designs to approximate the cost-to-go function and the optimal control policy at the same time \cite{Werbos-1989}. Such design approach has shown great promise to ameliorate the curse of dimensionality in dynamic programming. Among HDP algorithms, the dHDP \cite{Si-2001} is an online continual learning method that is intuitive and was motivated to be scalable and effective in solving realistic, large scale decision and control problems. Since its introduction, it has been applied to some complex control problems such as Apache helicopter stabilization and tracking control \cite{Enns-2000,Enns-2003}, damping low frequency oscillation in large power systems \cite{Lu-2008}, and most recently, robotic prosthesis control tested on amputee subjects \cite{Wen-2019}. However as previously described, we also noticed parameter drift due to continuous updates of learned neural networks \cite{Wen-2019,Gao-2020}. To avoid learning from undesired sensor and actuator noise as well as external disturbance, the control designs included a termination condition on the action and the critic network after sufficient learning of the control tasks during real life experimentation.

It is therefore desirable to have a systematic RL control method that would only update a prediction model by significant learning events but not by uncertain events or noise. And preferably, such learning should not (significantly) compromise the performance of the learned model. In this paper, we aim to develop an easy-to-implement event-based dHDP to update the policy and critic network weights driven by events directly reflected in system states. 

Our proposed work differs from and introduces new results beyond \cite{Zhong2-2016,Yang-2017,Dong-2017,Zhu-2017,Zhang-2019,Zhang-2016,Wang-2016,Wang-2017,Zhang2-2016,Yang-2018,Luo-2019}. Almost all previous event-triggered ADP algorithms were developed for continuous-time systems including [12]-[22], which also only deal with nonlinear dynamics that can be described as affine systems. Among the few existing works for discrete-time systems,  Sahoo $et \ al.$ \cite{Sahoo-2016} proposed a near optimal event-driven control for nonlinear systems using an ADP structure under some assumptions. However, this method was again, only applicable for affine systems and an identifier neural network was also necessary. More recently,  Ha $et \ al.$ \cite{Ha-2018} investigated an event based controller for affine discrete-time systems with constrained inputs. The stability of the systems was guaranteed with Lyapunov analysis tools but the approximation errors of the neural networks were ignored. Additionally, a third model neural network was needed. In [34], the authors proved the convergence of the weights for general discrete-time systems in an input to state stable framework but a model neural network was still needed. 

In this paper, our event-driven dHDP algorithm and its analysis represents new contributions to the important problem of event-driven ADP in the following aspects.

\begin{itemize}
	\item Compared with previous works \cite{Zhong2-2016,Yang-2017,Dong-2017,Zhu-2017,Zhang-2019,Zhang-2016,Wang-2016,Wang-2017,Zhang2-2016,Yang-2018,Luo-2019,Sahoo-2016,Ha-2018}, which require the nonlinear dynamics under consideration to be described as affine systems, our event-driven dHDP is applicable to general, discrete-time nonlinear systems.
	\item Compared with \cite{Sahoo-2016,Ha-2018,Dong3-2017} which require an identifier neural network in the controller design, our event-driven dHDP directly learns from data without the requirement of learning a dynamic system model either online or offline. Consequently the algorithm is easy to implement. Our previously demonstrated successes of time-driven dHDP in several important applications \cite{Enns-2000,Enns-2003,Lu-2008,Wen-2019,Gao-2020} speaks to the importance of an easy-to-implement algorithm.
	\item In addition to introducing a new, event-driven dHDP algorithm in this paper, we show the UUB of the neural network weights, the approximate solution to the Bellman equation approaching that of the Bellman optimality equation, and the stability of the controlled system. Our work is one of the few works that consider discrete-time nonlinear systems. Compared to those existing works addressing discrete-time nonlinear systems \cite{Sahoo-2016,Ha-2018,Dong3-2017}, we require fewer and/or milder assumptions. As such, our algorithm provides additional ease of implementation in applications and makes our results less conservative.
\end{itemize}

%
%

\section{Problem Formulation}

Consider a general nonlinear discrete-time system with unknown dynamics
\begin{equation}
x(k+1)=f(x(k), u(k)), 
\end{equation}
where $x\in \mathbb{R}^m$ is the system state and $u \in \mathbb{R}^n$ system input.

\textbf{Assumption 1}: System (1) is controllable. $x(k)=0$ is a unique equilibrium point and $f(0,0)=0$.

In time-driven ADP, a control input $u(k)$ is generated at each sampling time $k$ as a function of system state $x(k)$. In event-driven ADP, which is considered in this paper, the control input is updated only when there is a driving event reflected in system states going beyond a threshold. As such, the control inputs will be kept constant in a zero-order holder (ZOH) after a driving event time instant. We define the event indices as $\{\delta_k\}^{\infty}_{k=0}$ $(\delta_0=0)$. The control law can then be represented as
{\setlength\abovedisplayskip{1pt}
\setlength\belowdisplayskip{1pt}
\begin{equation}
u(k)=u(\delta_k), \forall \delta_k \leqslant k < \delta_{k+1}.
\end{equation}
}
We introduce the event-driven state error as
\begin{equation}
e(k)=x(\delta_k)-x(k), \forall \delta_k \leqslant k < \delta_{k+1}.
\end{equation}
Then system (1) can be rewritten as 
\begin{equation}
x(k+1)=f(x(k), u(e(k)+x(k)). 
\end{equation}
Let the event instants be determined by the following condition 
\begin{equation}
\parallel e(k) \parallel \leqslant e_T,
\end{equation}
where $e_T$ is a time varying threshold variable that will be investigated in next section.

Consider the following cost-to-go function with event-driven control
\vspace{-0.3cm}
\begin{equation}
\setlength{\belowdisplayskip}{-0.1cm}
V(x(k)) = \sum_{j=k}^{\infty} r(x(j), u(x(j)+e(j))),
\end{equation}
where 
\begin{equation}
\begin{array}{l}
r(x(k), u(x(k)+e(k))=r(x(k), u(x(\delta_k)) \\ 
=x^T(k)Qx(k)+u^T(x(\delta_k))Ru(x(\delta_k)).
\end{array}
\end{equation}

In the above equation, both $Q$ and $R$ are positive definite matrices whose maximum eigenvalue and minimum eigenvalue are $\lambda_{max}(\cdot)$ and $\lambda_{min}(\cdot)$, respectively. For simplicity, we denote $r(x(k), u(x(\delta_k))$ as $r(k)$.

Note that the $V(x(k))$ reflects a measure of the performance index at state $x(k)$ and satisfies the Bellman equation
\begin{equation}
	V(x(k))=r(k)+V(x(k+1)).
\end{equation}

From Bellman's optimality principle, the optimal cost-to-go function is therefore
{\setlength\abovedisplayskip{2pt}
	\setlength\belowdisplayskip{1pt}
\begin{equation}
	V^*(x(k)) = \min \limits_{u(x(\delta_k))} \{ r(k)+V^{*}(x(k+1)) \}.
\end{equation}
}

\vspace{-0.3cm}
\begin{figure}[thpb]
	\centering
	\includegraphics[scale=0.22]{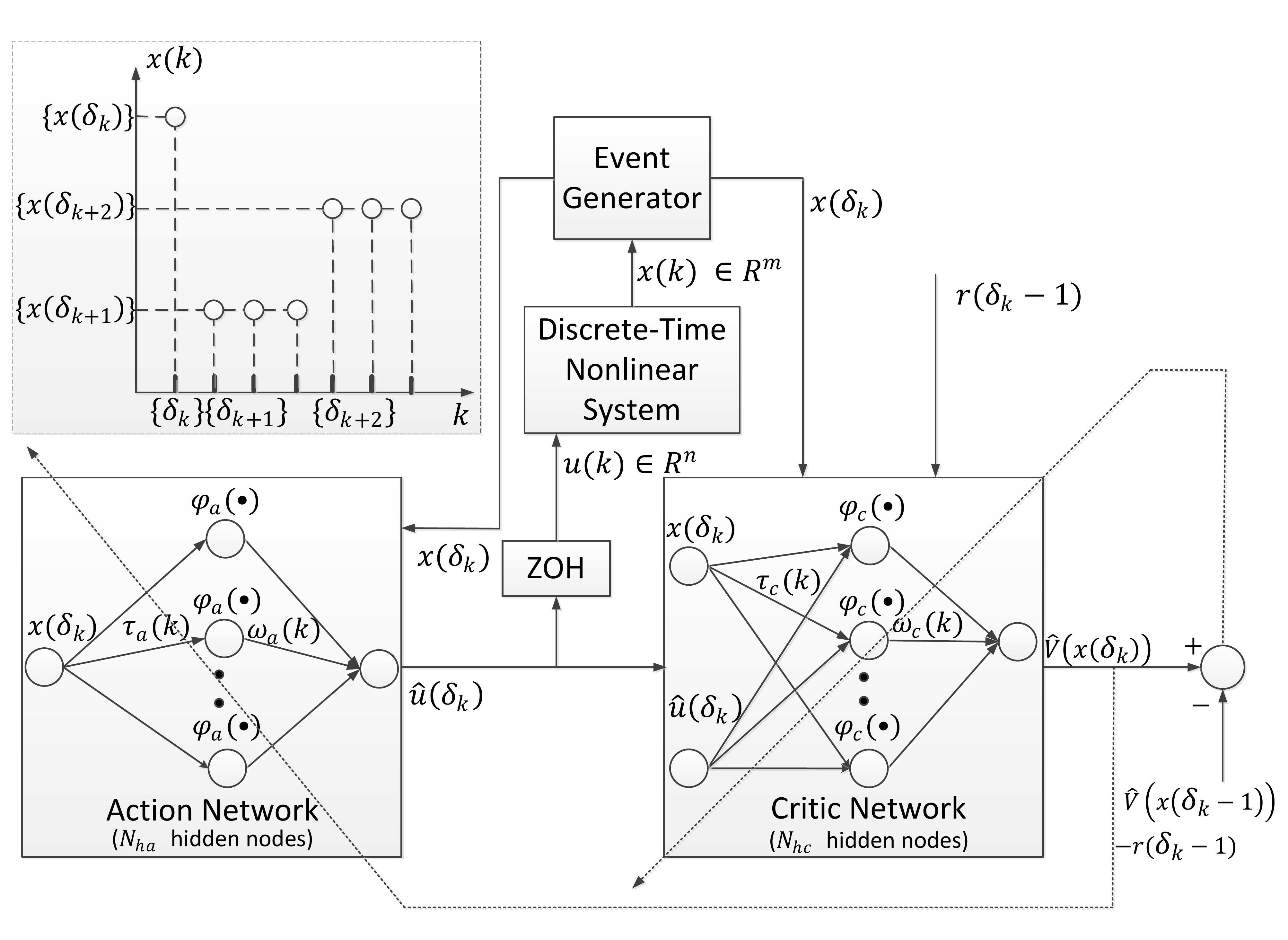}
	\caption{The relationship between event time instants $\delta_k$ and general time $k$ is shown in the upper left corner. As Eq. (2) and Eq. (3) show, only at event time instants $\delta_k$, the state $x(k)=x(\delta_k)$ is used to update $u(\delta_k)$. At other time instants, the ZOH element provides the needed control input to interact with the system. 
  	Correspondingly as shown in Eq. (14) and Eq. (17), the weights in the critic and action networks learn at event time instants $k=\delta_k$ and remain unchanged without triggering event. The critic network approximates the cost-to-go as a function of $x(\delta_k)$ and $u(\delta_k)$, while the action network approximates the input control as a function of $x(\delta_k)$.}
	\label{figure1}
\end{figure}
\vspace{-0.3cm}

To achieve event-driven learning control by dHDP, we adopt the architecture as shown in Fig. 1, where both the critic network and the action network are universal approximating neural networks. The nonlinear thresholding functions $\phi_c(.)$ and $\phi_a(.)$ in the hidden layer are hyperbolic tangents. We use $\tau_c(k), \tau_a(k)$ and $\omega_c(k), \omega_a(k)$ to denote the input-to-hidden layer weights and the hidden-to-output layer weights for the critic and the action networks, respectively. 

Given Fig. 1, we consider the approximate cost-to-go function $\hat{V}(x(k))$ and the approximate input $\hat{u}(k)$ of the following form, 
{\setlength\abovedisplayskip{1pt}
	\setlength\belowdisplayskip{1pt}
\begin{flalign}
& &\hat{V}(x(k))=\hat{\omega}^T_c(k)\phi_c(k),& &
\\
& &\hat{u}(k)=\hat{\omega}^T_a(k)\phi_a(k).& &
\end{flalign}
}

In this paper, the input-to-hidden layer weights $\tau_c(k), \tau_a(k)$ are chosen initially at random and kept constants as was the case in \cite{Liu-2012} and only the output layer weights $\omega_c(k), \omega_a(k)$ are updated during learning. As shown in \cite{Liu-2012}, such neural networks can be universally approximating.  

The critic network weights are updated in order to minimize the following approximation error, 
\begin{equation}
 E_c(k)  = \dfrac{e^T_c(k)e_c(k)}{2},  
\end{equation}
\begin{equation}
 e_c(k)  = \hat{V}(x(k))-[\hat{V}(x(k-1)-r(k-1)]. 
\end{equation}
Under the event-driven learning mechanism (2), learning and adaptation takes place only at event instants, i.e., $k=\delta_k$ and the control input holds as a constant during the event intervals. Then the critic network weights are updated as
\begin{numcases}{\hat{\omega}_c(k+1)=}
\setlength{\abovedisplayskip}{-0.6cm} 
\setlength{\belowdisplayskip}{-0.3cm}		
\hat{\omega}_c(k)-l_c\phi_c(k)[\hat{\omega}^T_c(k)\phi_c(k)+r(k-1)\notag\\\notag
-\hat{\omega}^T_c(k-1)\phi_c(k-1)]^T, \ \ k=\delta_k \\
\hat{\omega}_c(k), \qquad\qquad\qquad \,  \delta_k<k<\delta_{k+1},
\end{numcases}
where $l_c$ denotes the learning rate of the critic network.

Similar to \cite{Si-2001}, the action network weights are adjusted to minimize the following approximation error:
\begin{flalign}
& &E_a(k)=\dfrac{e^T_a(k)e_a(k)}{2},& &
\\
& &e_a(k)=\hat{V}(x(k)).& &
\end{flalign}
Similar to the critic network, the action network weights are updated as follows considering that the weight updates are driven by significant events in the system states.
\begin{numcases}{\hat{\omega}_a(k+1)=}
\setlength{\abovedisplayskip}{-0.3cm} 
\setlength{\belowdisplayskip}{-0.3cm}		
\notag \hat{\omega}_a(k)-l_a\phi_a(k)[\hat{\omega}^T_c(k)C(k)]\\
\times[\hat{\omega}^T_c(k)\phi_c(k)]^T, \qquad \ \ k=\delta_k \notag\\
\hat{\omega}_a(k), \qquad\qquad\quad   \delta_k<k<\delta_{k+1},
\end{numcases}
where $l_a$ denotes the learning rate of the action network and the components of $C(k)\in \mathbb{R}^{N_{hc}\times n} $ are expressed as
\begin{equation}
\begin{array}{l}
C_{li}(k)=\dfrac{1}{2}(1-\phi_{c_l}^2(k))\tau_{c_{l,m+i}}, \\ 
l=1, \cdots, N_{hc}, i=1, \cdots, n,
\end{array}
\end{equation}
where $l$ and $i$ are the indices of the hidden and input neurons, while $m$ and $n$ denote the dimension of system state and input.

We refer the critic and action update rules in (14) and (17) as event-driven dHDP. Let $\omega_c^*$ and $\omega_a^*$ be the optimal weights of the critic network and the action network, respectively, i.e., 
\begin{equation}
\omega_c^*=\arg \min \limits_{\omega_c}|| \hat{V}(x(k))-[\hat{V}(x(k-1)-r(k-1)]||,
\end{equation}
\begin{equation}
\setlength{\abovedisplayskip}{-0.3cm} 
\setlength{\belowdisplayskip}{-0.3cm}
\omega_a^*=\arg \min \limits_{\omega_a} \parallel \hat{V}(x(k)) \parallel.
\end{equation}
Then we have 
\begin{flalign}
& &V^*(x(k))=\omega^{*T}_c(k)\phi_c(k)+\epsilon_c,& &
\\
& &u^*(k)=\omega^{*T}_a(k)\phi_a(k)+\epsilon_a,& &
\end{flalign}
where $V^*(x(k))$ and $u^*(k)$ denote the optimal cost-to-go function and optimal control input, while $\epsilon_c$ and $\epsilon_a$ are the neural network reconstruction errors of the critic network and the action network, respectively.

\section{Main Results}
As previously described, our goal is to devise a dHDP-based ADP online learning method that adapts the action and the critic weights during significant system events reflected in system states. That is to say that learning and adaptation does not necessarily take place regularly at each sampling time when an observation is made. We therefore will provide a sufficient condition under which learning takes place as driven by events while the closed-loop system stability is guaranteed. Specifically we will show that both the weight parameters and the system states will remain UUB.  

We define the weight estimation errors and some auxiliary variables as follows:

\begin{equation}
\setlength{\abovedisplayskip}{-0.2cm} 
\setlength{\belowdisplayskip}{6pt}
\begin{array}{l}
\tilde{\omega}_c(k)=\hat{\omega}_c(k)-\omega_c^*, \\ 
\xi_c(k)=(\hat{\omega}_c(k)-\omega_c^*)^T\phi_c(k)=\tilde{\omega}^T_c(k)\phi_c(k),\\
\tilde{\omega}_a(k)=\hat{\omega}_a(k)-\omega_a^*, \\ 
\xi_a(k)=(\hat{\omega}_a(k)-\omega_a^*)^T\phi_a(k)=\tilde{\omega}^T_a(k)\phi_a(k).
\end{array}
\end{equation}

\textbf{Assumption 2}: For the critic network and the action network, the optimal weights, the activation functions and the reconstruction errors are bounded, i.e., $\parallel \omega_c^* \parallel \leqslant \omega_{cm}$, $\parallel \omega_a^* \parallel \leqslant \omega_{am}$, $\parallel \phi_c \parallel \leqslant \phi_{cm}$,$\parallel \phi_a \parallel \leqslant \phi_{am}$, $\parallel \epsilon_c \parallel \leqslant \epsilon_{cm}$, $\parallel \epsilon_a \parallel \leqslant \epsilon_{am}$. Besides, the activation function $\phi_a$ is Lipschitz continuous and satisfies 
\begin{equation}
\parallel \phi_a (x_1)  -  \phi_a (x_2) \parallel\leqslant L_a\parallel x_1-x_2 \parallel,
\end{equation}
for all $x_1 ,x_2 \in \Omega$ where $L_a$
is a positive constant and $\Omega$ is the domain of function $f(x,u)$.

\textbf{Theorem 1}: Consider nonlinear system (1) and let Assumptions 1-2 hold. Let the critic and action network weights update iteratively according to (14) and (17) if a learning event occurs based on the following event-driven state error criterion (25),  
\begin{equation}
\parallel e(k) \parallel^2 \leqslant \dfrac{\lambda_{min}(Q)\beta}{2\lambda_{max}(R)\parallel\hat{\omega}_a \parallel^2 L^2_a}\parallel x(k) \parallel^2,
\end{equation}
and let the learning rates of the action and critic networks satisfy 
\begin{equation}
l_c < \frac{1}{\parallel \phi_c(k) \parallel^2}, l_a < \frac{1}{\parallel \phi_a(k) \parallel^2},
\end{equation}
then we have the following results:

\begin{itemize}
	\item[1.]  The errors between the optimal network weights $\omega_c^*$, $\omega_a^*$ and their estimates $\hat{\omega}_c(k)$ and $\hat{\omega}_a(k)$ are UUB. 
	\item[2.] The control $u(k)$, which is parameterized by $\omega_a(k)$, is a stabilizing control to guarantee the UUB of the nonlinear system under the proposed event-driven dHDP algorithm.
\end{itemize}

\begin{IEEEproof}
	We consider the following two cases.
	
	$Case 1$: No learning event is observed at time instant $k$. Consider the Lyapunov function candidate defined as follows:
	\begin{equation}
	\setlength{\abovedisplayskip}{3pt} 
	\setlength{\belowdisplayskip}{3pt}
	L(k)=L_1(k)+L_2(k)+L_3(k)+L_4(k),
	\end{equation}
	where 
	\begin{align*}
	\setlength{\abovedisplayskip}{3pt} 
	\setlength{\belowdisplayskip}{3pt}
	L_1(k)=\dfrac{1}{l_c} {\rm tr} [\tilde{\omega}^T_c(k)\tilde{\omega}_c(k)],
&
	\quad L_2(k)=\dfrac{1}{\gamma l_a} {\rm tr} [\tilde{\omega}^T_a(k)\tilde{\omega}_a(k)],\\
	L_3(k)= \frac{1}{2}\parallel \xi_c(k-1) \parallel^2,
&
	\qquad L_4(k)=V(x(k)).
	\end{align*}
	As $\tilde{\omega}_c(k+1)=\tilde{\omega}_c(k)$ and $\tilde{\omega}_a(k+1)=\tilde{\omega}_a(k)$, we find $\triangle L_1(k)= \triangle L_2(k)=0$.
	\vspace{-0.2cm}
	\begin{flalign}
	&\, \triangle L_3(k)= \frac{1}{2}[\parallel \xi_c(k) \parallel^2-\parallel \xi_c(k-1) \parallel^2],&\\\notag
	&\, \triangle L_4(k)= V(x(k+1))-V(x(k))=-r(k)&\\\notag
	&=-x^T(k)Qx(k)-u^T(x(\delta_k))Ru(x(\delta_k))&\\\notag
	&=-x^T(k)Qx(k)-\{-[u^*(x(k))-u(x(\delta_k))]&\\\notag
	&+u^*(x(k))\}^TR \{-[u^*(x(k))-u(x(\delta_k))]+u^*(x(k))\}&\\\notag
	&=-x^T(k)Qx(k)-u^{*T}(x(k))Ru^*(x(k))&\\\notag
	&-[u^*(x(k))-u(x(\delta_k))]^TR[u^*(x(k))-u(x(\delta_k))]&\\
	&+2u^{*T}(x(k))R[u^*(x(k))-u(x(\delta_k))].&
	\end{flalign}
	From (22), we have
	\begin{flalign}
	&u^*(x(k))\!-\!u(x(\delta_k))\!=\!\omega^{*T}_a(k)\phi_a(k)\!+\!\epsilon_a\!-\!\hat{\omega}^T_a(k)\phi_a(x(\delta_k))&\notag\\ &=\!\hat{\omega}_a^T(k)[\phi_a(x(k))\!-\!\phi_a(x(\delta_k))]-\xi_a(k)+\epsilon_a.&
	\end{flalign}
	Substituting (24) and (30) into (29), we obtain
	\begin{flalign}
	&\, \triangle L_4(k) \leqslant -x^T(k)Qx(k)+\lambda_{max}(R)\parallel u^*(x(k)) \parallel^2&\notag\\\notag
	&+\lambda_{max}(R)[\hat{\omega}_a^T(k)[\phi_a(x(k))-\phi_a(x(\delta_k))]&\\\notag
	&-\tilde{\omega}^T_a(k)\phi_a(x(k))+\epsilon_a]^2&\\\notag
	&\leqslant -\lambda_{max}(Q)\parallel x(k) \parallel^2+2\lambda_{max}(R)\parallel\hat{\omega}_a \parallel^2 L^2_a \parallel e(k) \parallel^2 &\\
	&+6\lambda_{max}(R)(\omega_{am}^2\phi_{am}^2+\epsilon_{am}^2).&
	\end{flalign}
	Recall (25), from (28) and (31), we have
	\begin{equation}
	\triangle L(k) \leqslant -(1-\beta)\lambda_{min}(Q) \parallel x(k) \parallel^2+D_{1m}^2,
	\end{equation}
	where 
	\begin{equation}
	D_{1m}^2=6\lambda_{max}(R)(\omega_{am}^2\phi_{am}^2+\epsilon_{am}^2)+2\omega_{cm}^2\phi_{cm}^2.
	\end{equation}	
	If 
	\begin{equation}
	\parallel x(k) \parallel >\dfrac{D_{1m}}{\sqrt{(1-\beta)\lambda_{min}(Q)}},
	\end{equation}
	the first difference $\triangle L(k) \le 0$. This demonstrates that the closed-loop system state and the weight estimation errors are UUB in this case. 
	
	$Case 2$: A learning event is observed at time instant $k$ with an event index $\delta_k$. In this case, the weights of the critic network and the action network will be updated using (14) and (17). With the same Lyapunov function candidate (27) and a similar derivation as Theorem 4.3 in \cite{Liu-2012}, we have
	\begin{flalign}
	&\triangle L(k) \leqslant -(\dfrac{1}{2}\!-\!\dfrac{4}{\gamma})|| \xi_c(k)||^2-\lambda_{max}(Q)|| x(k)||^2+D_{2m}^2, &\notag\\
	&D_{2m}^2=(12\!+\!\dfrac{4}{\gamma})\omega_{cm}^2\phi_{cm}^2\!+ \dfrac{4}{\gamma} \omega_{cm}^2C_m^2\omega_{am}^2\phi_{am}^2+8r_m^2&
	\end{flalign}
	where $C_m$ and $r_m$ are the upper bounds of $C(k)$ and $r(k)$ in (18) and (7), respectively, and $\gamma > 8$ is a weighting factor. If
	\begin{align}
	\parallel x(k) \parallel >\dfrac{D_{2m}}{\sqrt{\lambda_{min}(Q)}} \quad \rm{or}
&
	\ \parallel \xi_c(k) \parallel >\dfrac{D_{2m}}{\sqrt{(\dfrac{1}{2}-\dfrac{4}{\gamma})}},
	\end{align}
	the first difference $\triangle L(k) \le 0$.

	From Case 1 and Case 2, we conclude that the closed-loop system state and the errors between the optimal network weights $\omega_c^*$, $\omega_a^*$ and their respective estimates $\hat{\omega}_c(k)$ and $\hat{\omega}_a(k)$ are UUB. This completes the proof.
\end{IEEEproof}

\textbf{Remark 1}: Theorem 1 presents a sufficient condition to guarantee the stability of an event-driven dHDP controlled system with learning realized in the critic and action neural networks. Without any special constraints on the nonlinear system, we have obtained a system boundedness result under a few mild assumptions. As there exist unavoidable approximation errors in both the cost-to-go function and the input control, Theorem 2 below examines how an approximate solution of the Bellman optimality equation can be achieved within a finite approximation error.

\textbf{Theorem 2}: Under the same conditions as in Theorem 1, the Bellman optimality equation is approximated within a finite approximation error. Meanwhile, the adopted control law $\hat{u}(k)$ is uniformly convergent to a finite neighborhood of the optimal control $u^*(k)$.

\begin{IEEEproof}
	From the approximate cost-to-go in equation (10) for the Bellman equation
	(8) and the optimal cost-to-go expressed in approximation form (21)  
	for the Bellman optimality equation (9), we have
	\begin{flalign}
	\setlength{\abovedisplayskip}{1pt} 
	\setlength{\belowdisplayskip}{1pt}
	&\parallel \hat{V}(x(k)) - V^*(x(k)) \parallel = \parallel \hat{\omega}^T_c(k)\phi_c(k) - \omega^{*T}_c(k)\phi_c(k)&\notag \\
	&-\epsilon_c \parallel \leqslant \parallel \tilde{\omega}_c(k) \parallel \phi_{cm}  + \epsilon_c \leqslant \hat{\epsilon}_c.& 
	\end{flalign}

	Similarly, from (11) and (22), we have
	\begin{equation}
	\setlength{\abovedisplayskip}{3pt} 
	\setlength{\belowdisplayskip}{3pt}
	\parallel \hat{u}(k) - u^*(k) \parallel \leqslant \parallel \tilde{\omega}_a(k) \parallel \phi_{am}  + \epsilon_a \leqslant \hat{\epsilon}_a.
	\end{equation}
	This comes directly as $\parallel \tilde{\omega}_c(k) \parallel$ and $\parallel \tilde{\omega}_a(k) \parallel$ are both UUB as shown in Theorem 1. This demonstrates that the Bellman optimality is achieved within finite approximation errors. 
\end{IEEEproof}

\section{An Illustrative Example}  
In this section, we use a numerical example to demonstrate the theoretical results in this paper and differences between event-driven dHDP and its time-driven counter part. The unknown system dynamics are assumed to be generated from the following equation,
\begin{equation*}
\setlength{\abovedisplayskip}{3pt} 
\setlength{\belowdisplayskip}{3pt}
\left[
\begin{matrix}
x_1(k+1) \\
x_2(k+1) \\
\end{matrix}\right]
=\left[
\begin{matrix}
0.9996x_1(k)+0.0099x_2(k) \\
-0.0887x_1(k)+0.99x_2(k) \\
\end{matrix}
\right] +\left[
\begin{matrix}
0 \\
0.1u(k) \\
\end{matrix}\right].
\end{equation*}

Now we try to stabilize this system by the proposed event-driven dHDP algorithm. For both the critic neural network and the action neural network, we set the number of hidden layer nodes to $N_{hc}=N_{ha}=6$. From Theorem 1, the learning rates of these two neural networks should satisfy equation (26). In this simulation, we choose $l_a=l_c=0.1$. The initial condition of the system state is set as $x(0)=[-1,1]^T$. The RL signal is $r(k)=x^T(k)I_mx(k)+u^T(k)(0.1I_n)u(k)$, where $I_m$ and $I_n$ are identity matrices. The initial weights of both the critic and the action neural network are set randomly within $[-0.4, 0.4]$. For the learning event condition (25), we choose different $\beta$ values where $\beta = 0$ corresponds to time-driven dHDP as in \cite{Si-2001}.

Dynamic system state trajectories under different event-driven learning conditions (i.e., different $\beta$ values) are shown in Fig. 2. As expected, little difference is noticed between the event-driven dHDP and the time-driven dHDP when $\beta$ is small.  Also as expected, the controlled system state trajectories deviate further from those controlled by the time-driven dHDP as $\beta$ increases. Fig. 3 illustrates reduced numbers of learning events 
and consequently degraded learning control responses as $\beta$ increases.

 \begin{figure}[thpb]
 	\centering
 	\includegraphics[height=5.2cm,width=8cm]{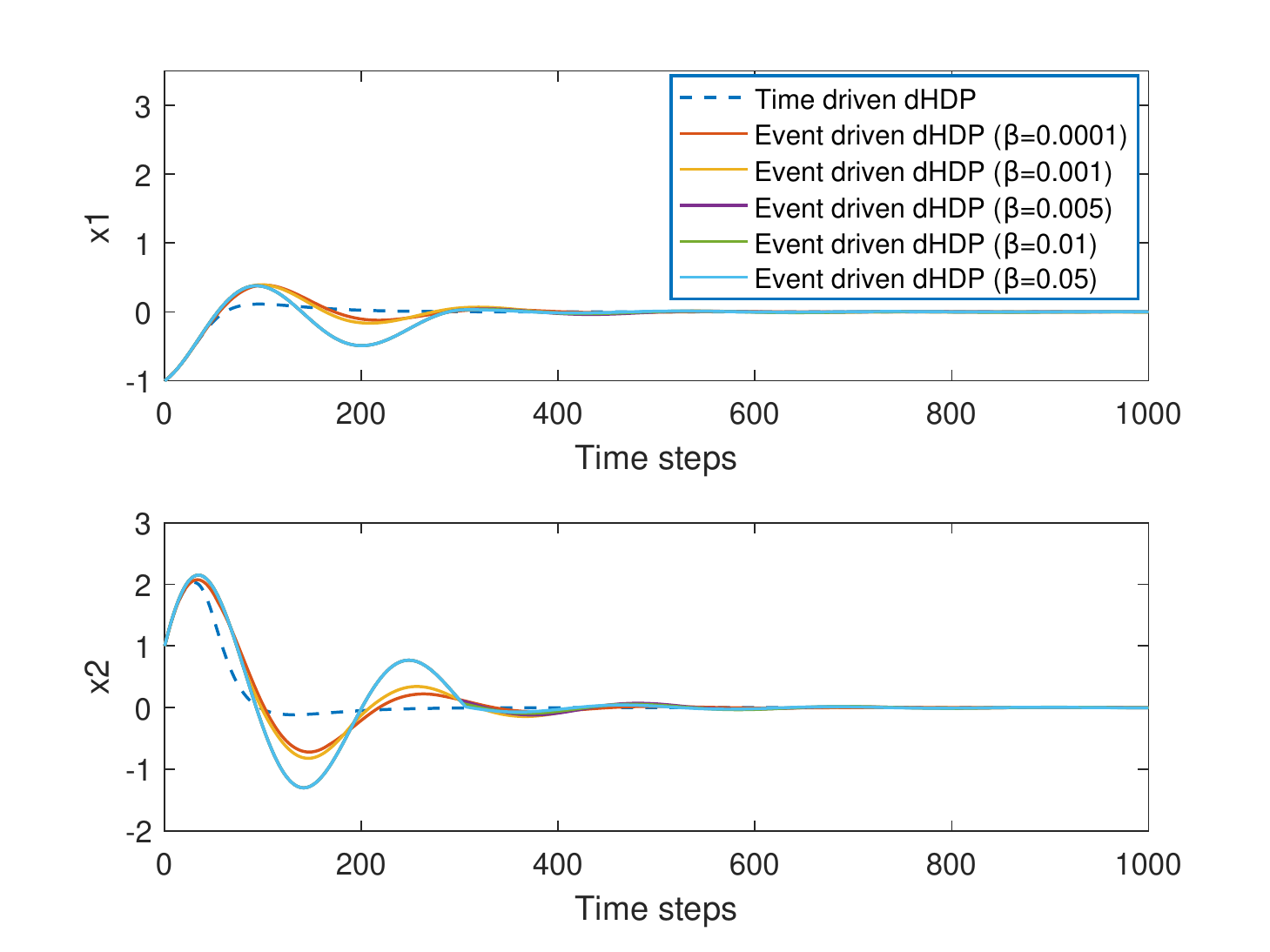}
 	\caption{State trajectories with event-driven dHDP controller (different $\beta$) and
 		time-driven dHDP controller ($\beta=0$).}
 	\label{figure2}
 \end{figure}
 \begin{figure}[thpb]
 	\centering
 	\includegraphics[height=5.2cm,width=8cm]{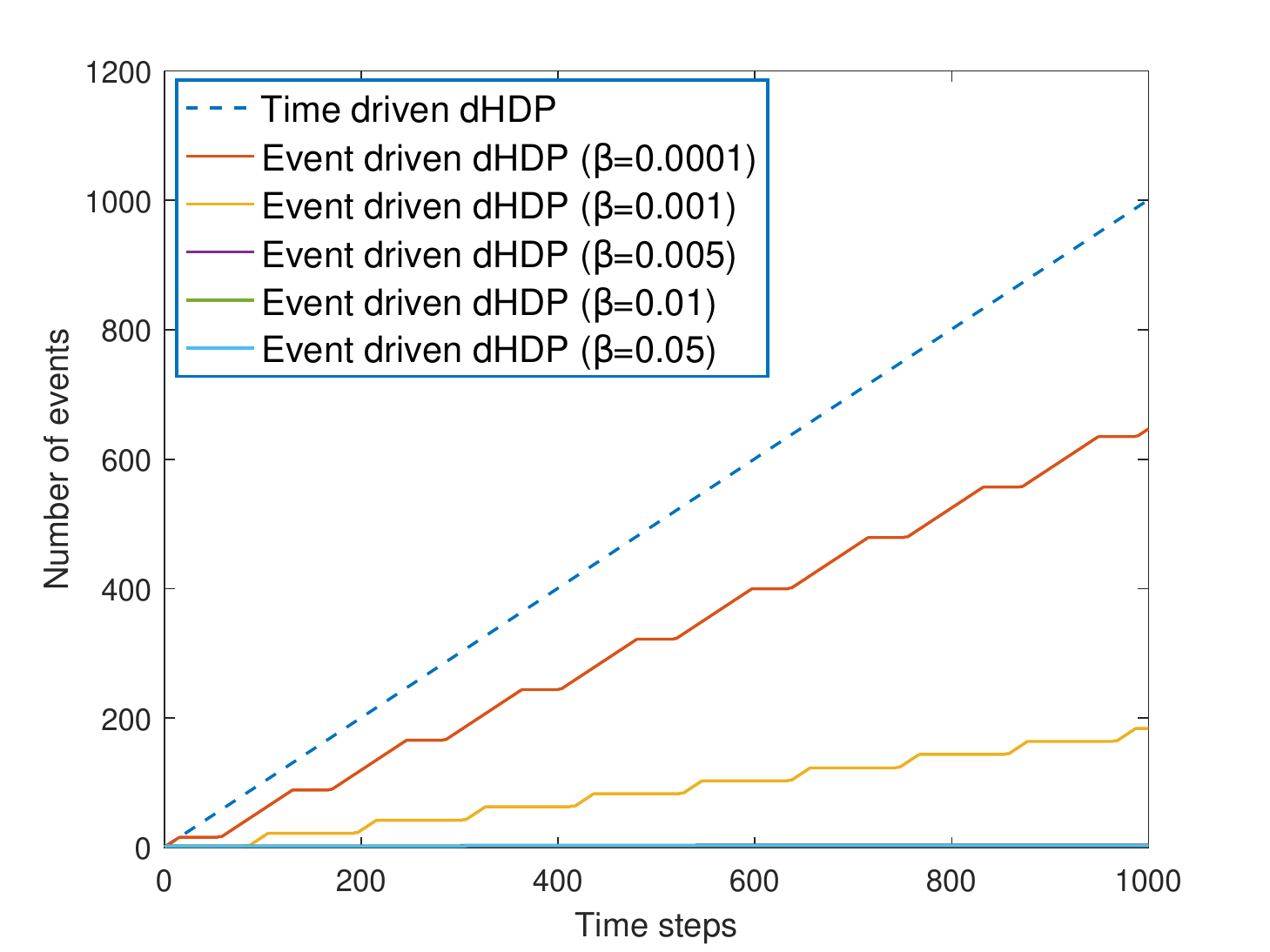}
 	\caption{Numbers of learning events of event-driven dHDP (different $\beta$) and time-driven dHDP ($\beta=0$).}
 	\label{figure3}
 \end{figure}
 \begin{figure}[thpb]
 	\centering
 	\includegraphics[height=5.2cm,width=8cm]{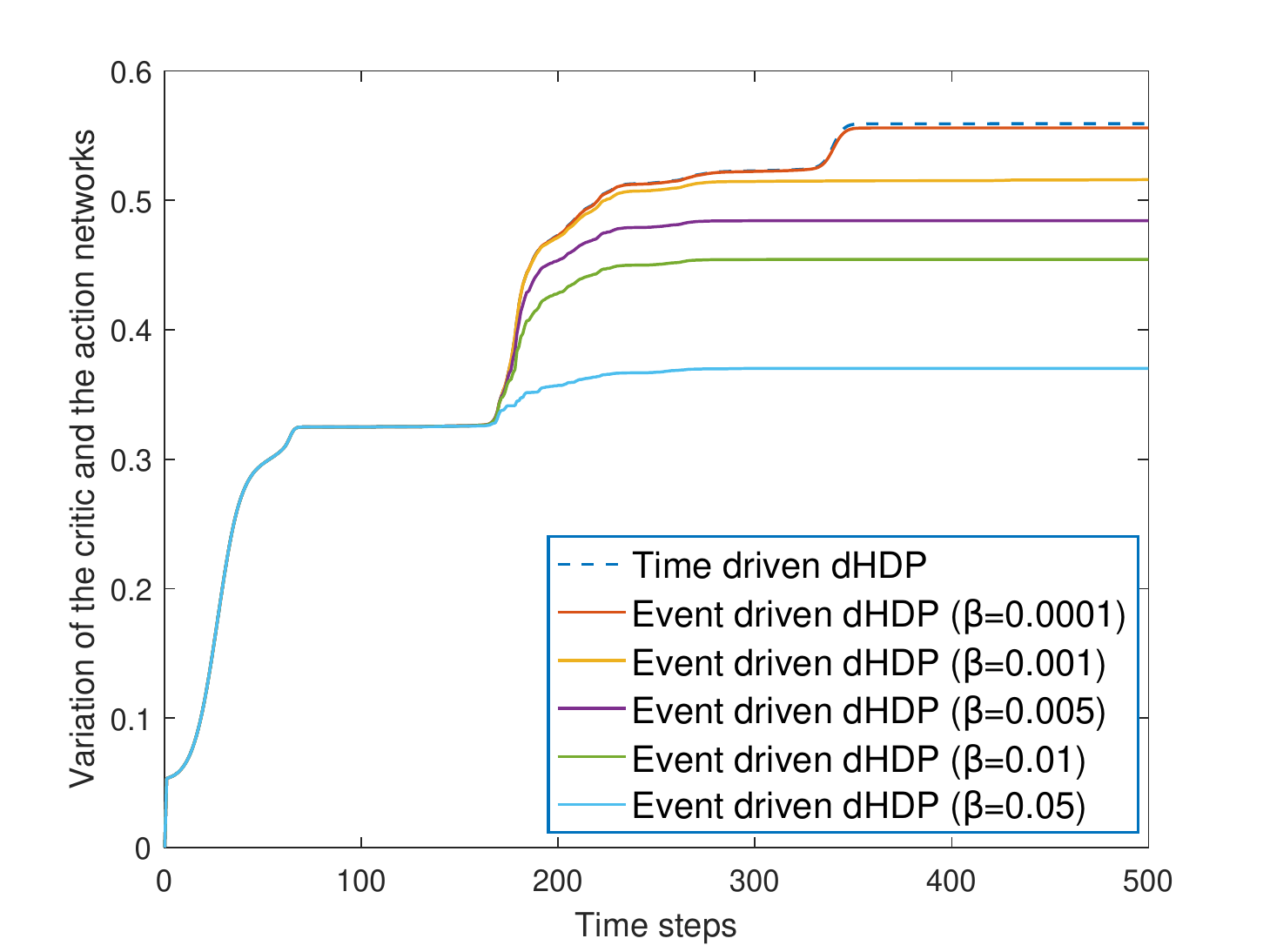}
 	\caption{Average accumulated variation for every weight under event-driven dHDP controller (different $\beta$) and time-driven dHDP controller ($\beta=0$) with external noise.}
 	\label{figure4}
 \end{figure}
 
Next, we illustrate how event-driven dHDP may be used as an effective tool to prevent converged critic and action network weights from drifting too far from the learned controller due to inevitable noise in the state measurements.
 
This simulation entails 500 time steps, the first 100 of which are for training the critic and action weights. During this period, the weights are updated at each time instant $k$ while the system dynamics are not subject to any noise. From time step 101 to 300, we introduce random Gaussian white noise $\omega_k \sim N(\omega_0,1)$ with $\omega_0=0.1$ into the system. Then from time step 301 to 500, the random Gaussian white noise is removed or the system dynamics evolve under noise free condition.  We use event-driven dHDP with different $\beta$ values and consider the accumulated amount of variation in the critic and action network weights as defined below:
 \begin{equation}
 	\eta (k) = \dfrac{\sum_{i=a,c} \sum_{j=0}^{k} \parallel \hat{\omega}_i(j+1)-\hat{\omega}_i(j) \parallel^2}{N_{hc}+N_{ha}},
 \end{equation}
 where $\parallel \cdot \parallel$ denotes the 2-norm.
  
 The first 100 time steps in Fig. 4 shows actual learning in the critic and action networks and such learning converges at about time step 80. From steps 101 to 300, we notice from small to large weight variations corresponding to small to large $\beta$ values due to the added white Gaussian noise, and such variation is the largest when $\beta=0$ or when learning takes place continuously. Therefore we can see how learning event condition (25) helps protect the critic and action networks from drifting away due to random fluctuation in system dynamics.

\section{Conclusion}
We have proposed a new event-driven RL control method, i.e., event-driven dHDP based on the time-driven dHDP for a general discrete time nonlinear system. Specifically we introduced a learning event criterion that is directly related to system states such that a learned dHDP controller could be less affected by random fluctuations in the system dynamics. We proved the stability of the closed-loop system under event-driven dHDP control by using Lyapunov functions as well as the approximate optimality of the dHDP control solution. We demonstrated a significant reduction of weight updating times for event-driven learning from time-driven learning without sacrificing too much system performance.

\bibliographystyle{IEEEtran}
\bibliography{ref_mixed_initiative}

%

\appendices
\end{document}